\documentstyle[12pt]{article}

\begin{document}

\newcommand{\beq}{\begin{equation}}
\newcommand{\eeq}{\end{equation}}
\newcommand{\m}{\mbox{\boldmath $m$}}
\newcommand{\n}{\mbox{\boldmath $n$}}
\newcommand{\cl}{\mbox{\boldmath $l$}}
\newcommand{\k}{\mbox{\boldmath $k$}}
\newcommand{\p}{\mbox{\boldmath $p$}}

\def\a{\alpha'}
\def\R{\hat{R}_1}
\def\S{\hat{R}_2}
\def\T{\hat{R}_3}
\def\U{\hat{R}_4}

\begin{titlepage}

\begin{center}

\hfill hep-th/9709006

\vskip 1.8cm
{\bf Membrane, Four-Brane and Dual Coordinates \\
 in the M(atrix) Theory Compactified on Tori}

\vskip .8cm

Shijong Ryang

\vskip .8cm
{\em Department of Physics \\ Kyoto Prefectural University of Medicine \\
Taishogun, Kyoto 603 Japan}

\end{center}

\vskip 2.0cm

\begin{center} {\bf Abstract} \end{center}
In the M(atrix) theory by making the expansions of the matrices around the
membrane and four-brane solutions we derive the three- and five-dimensional
gauge theories on the dual tori. The explicit forms of solutions yield
the dual coordinates and each expansion is related to a toroidal
compactification of the M(atrix) theory. From the derived Lorentz and
gauge invariant actions the gauge coupling constants are shown to be
characterized by the volume of the dual tori.

\vskip 7cm
\noindent August, 1997
\end{titlepage}

The M(atrix) theory has been proposed as a nonperturbative microscopic
description of M-theory \cite{BFSS}. The dynamics of M-theory in the infinite
momentum frame is described by the large N $10$-dimensional U(N) Super Yang
Mills theory (SYM) reduced to $0+1$ dimensions whose Hamiltonian consists
of D$0$-branes and their interactions. Its operator-like classical solutions
are identified with D$p$-branes of the type IIA superstring theory. The
long-distance interactions of some $p$-branes have been shown to be in
agreement with supergravity predictions \cite{AB,LM,GL}.

The M(atrix) theory description of toroidal compactification has been given
by Ganor, Ramgoolam and Taylor \cite{GRT,WT}, where the M(atrix) theory on
$T^3$ is T-duality equivalent to the U(N) SYM theory on the dual torus
$\hat{T}^3$ and T-duality is realized by S-duality in the $4$-dimensional
SYM theory. On the other hand starting
from the Lagrangian of the U(N) SYM theory in
$0+1$ dimensions and analyzing the small fluctuation around the general
classical $p$-brane solution Banks, Seiberg and Shenker have derived the
U(N) SYM theory in $p+1$ dimensions, which describes D$p$-brane in the
IIA string theory \cite{BSS}. In this derivation the notions of T-duality
as well as toroidal compactification are absent. We will fill up
the gap between the
two approaches. By using the explicit infinite membrane and four-brane
solutions and extracting dual variables from them we will derive the
gauge invariant actions on the dual spacetimes.

We start with the Lagrangian of the M(atrix) theory described by the U(N)
supersymmetric quantum mechanics $SYM_{0+1}$ \cite{BFSS}. The bosonic part
of it is given by
\beq
L = \frac{1}{2g_s}Tr \left( \frac{1}{\a^2}D_0X^iD_0X^i +
\frac{1}{2\a^4}[X^i,X^j]^2 \right),
\label{Lg}\eeq
where $D_0 = \partial_{t_0} - i[A_0,\cdot]$ and $i=1,2,\cdots,9$ \cite{D}.
The string coupling constant $g_s$ and $X^i$ have mass dimensions $3$
and $-1$ respectively. Through the rescalings such as $Y = X/g^{1/3},
t = t_0/g^{1/3}$ and $g_s = g(\a)^{-3/2}$ this Lagrangian can be
rewritten as
\beq
L = Tr \left( \frac{1}{2R}D_tY^iD_tY^i + \frac{R}{4\a^3}[Y^i,Y^j]^2
 \right),
\label{LR}\eeq
where $R = g^{2/3}\sqrt{\a}$.

We take an infinite membrane background stretched out in the $1, 2$ planes
to be
\begin{eqnarray}
U_1 = -\sqrt{N}R_1p,& &U_2 = \sqrt{N}R_2q
\label{U}\end{eqnarray}
with all others $U_I = 0, I = 3,\cdots,9$ where $R_r, r = 1,2$ are the
lengths of the corresponding directions and $[q,p] = 2{\pi}i/N$
\cite{BFSS,AB}. The minus sign of $U_1$ is suggested by the construction of
D-string action \cite{ML} from the D-instanton matrix model \cite{IKKT}.
Following the general framework of Ref.\cite{BSS}, we expand the matrix
variables around the explicit membrane background
\begin{eqnarray}
Y^1 = U_1 + 2{\pi}{\a}A_1,&  Y^2 = U_2 + 2{\pi}{\a}A_2,&  Y^I = Z_I.
\label{Y}\end{eqnarray}
Here we define dual matrices $x_1, x_2$ as
\begin{eqnarray}
x_1 = \frac{{\a}\sqrt{N}}{R_1}q, & & x_2 = \frac{{\a}\sqrt{N}}{R_2}p
\end{eqnarray}
and calculate commutation relations to have
\[[Y^1,Y^2]  =  2\pi iR_1R_2 + (2\pi\a)^2 iF_{12},  \]
\[D_tY^r  =  2\pi{\a}F_{tr}, \;r =1,2,  \]
\beq
[Y^r,Y^I]  =  2\pi{\a}iD_rZ_I,
\label{CR}\eeq
where $F_{12} = \partial_{x_1}A_2 - \partial_{x_2}A_1 -i[A_1,A_2]$.
From (\ref{Y}) and (\ref{CR}) the Lagrangian (\ref{LR}) can be
expanded as
\begin{eqnarray}
L &=& Tr ( \frac{(2\pi\a)^2}{2R}F_{tr}^2 -
\frac{(2\pi)^4R\a}{2}F_{12}^2 - \frac{2{\pi}^2R}{\a^3}(R_1R_2)^2
\nonumber \\  & & + \frac{1}{2R}(D_tZ_I)^2 - \frac{2\pi^2R}{\a}(D_rZ_I)^2
+ \frac{R}{4\a^3}[Z_I,Z_J]^2 ).
\label{LT}\end{eqnarray}
In this expansion the equations of motion for the static configuration
are taken into account by dropping linear terms in the fluctuations.

In the $A_t = 0$ gauge a Hamiltonian is provided by
\begin{eqnarray}
H &=& RTr ( \frac{1}{2(2\pi\a)^2}E_r^2 + \frac{(2\pi)^4\a}
{2}F_{12}^2 + \frac{1}{2}{\Pi}_I^2 \nonumber \\
 & & {}+ \frac{2\pi^2}{\a}(D_rZ_I)^2 + \frac{2\pi^2}{\a^3}(R_1R_2)^2
 - \frac{1}{4\a^3}[Z_I,Z_J]^2 ).
\label{HR}\end{eqnarray}
We will consider the membranes of toroidal topologies in the large N limit.
Using a pair of unitary operators $U$, $V$
satisfying $UV = \omega VU$,
$U^N = V^N = 1$ with $\omega = \exp(2\pi i/N)$ we have a set of matrices
$J_{\m} = {\omega}^{-m_1m_2/2}U^{m_1}V^{m_2}$ for $\m = (m_1,m_2)$ and
$0 \leq m_1,m_2 \leq N-1$, which include $J_{00}$ and are a linearly
independent, complete set of basis for U(N) matrices. The fluctuations
around the background and their conjugate momenta are expanded in
the form
\[
A_r = \frac{1}{2\pi\sqrt{\a}}\sum_{\m}A_{\m}^rJ_{\m}, \hspace{1cm}
E_r = 2\pi\sqrt{\a}\sum_{\m}E_{\m}^rJ_{\m}, \]
\beq
Z_I = \sqrt{\a}\sum_{\m}Z_{\m}^IJ_{\m}, \hspace{1cm}
 \Pi_I = \frac{1}{\sqrt{\a}}\sum_{\m}\Pi_{\m}^IJ_{\m}.
\label{AZ}\eeq
The matrices $J_{\m}$ satisfy an orthonormal relation
\beq
TrJ_{\m}J_{\n} = N \delta_{m_1+n_1,0}\delta_{m_2+n_2,0},
\label{J}\eeq
where $\delta$ is a mod.N delta function. Since $U$ and $V$ are realized
to be exponentials of the canonical variables $p, q$ as
$U = e^{ip}, V = e^{iq}$ the expressions (\ref{AZ})
are understood as the mode expansions
in the quantum phase space $(p,q)$. Using $p$ and $q$ leads to the
alternative expression $J_{\m}(p,q) = \exp[i(m_1p+m_2q)]$. By plugging the
mode expansions (\ref{AZ}) into the Hamiltonian (\ref{HR}) and taking
the trace by means of (\ref{J}) we obtain the decomposition
\begin{eqnarray}
H &=& R( H_0 + H_1 + H_2 ), \nonumber\\
H_0 &=& N \left( \frac{1}{2\a}\sum_{\m}(\Pi_{-\m}^I \Pi_{\m}^I +
E_{-\m}^r E_{\m}^r) + \frac{2}{\a^3}(\pi R_1R_2)^2 \right) \nonumber\\
 & &  + \frac{2\pi^2}{\a^2}\sum_{\m}
\left( (R_1m_2A_{-\m}^2-R_2m_1A_{-\m}^1)
(R_1m_2A_{\m}^2-R_2m_1A_{\m}^1) + \omega_{\m}^2Z_{-\m}^IZ_{\m}^I \right),
\nonumber\\
H_1 &=& \frac{4\pi i}{\sqrt{N}(\a)^{3/2}}\sum_{\m,\n,\cl}N\sin\frac{\pi}{N}
(\n \times \cl) ( (R_1m_2A_{\m}^2-R_2m_1A_{\m}^1)A_{\n}^1A_{\cl}^2
\nonumber\\
  & & \mbox{} + Z_{\m}^I(R_1m_2A_{\n}^1+R_2m_1A_{\n}^2)Z_{\cl}^I)
 \delta_{\m+\n+\cl,0}, \nonumber\\
 H_2 &=& \frac{1}{\a}\sum_{\m,\n,{\cl},\k}N \sin \frac{\pi}{N}(\m \times \n)
\sin \frac{\pi}{N}(\cl \times \k)( Z_{\m}^IZ_{\n}^JZ_{\cl}^IZ_{\k}^J
\nonumber\\ & & + 2A_{\m}^1 A_{\n}^2 A_{\cl}^1 A_{\k}^2 +
2A_{\m}^rZ_{\n}^IA_{\cl}^rZ_{\k}^I ) \delta_{\m+\n+\cl+\k,0}
\label{HRH}\end{eqnarray}
with $\omega_{\m} = \sqrt{(R_1m_2)^2 + (R_2m_1)^2}$ and $\m \times \n =
m_1n_2 - m_2n_1$. In the large N expansion we observe that the leading and
the next leading parts $H_0$ show the free quadratic behavior.
Specially the quadratic terms with respect to $(Z_{\m}^I, \Pi_{\m}^I)$
indicate the Hamiltonian of an infinite set of harmonic oscillators
characterized by $(m_1,m_2)$. The similar structure was observed in the
zero limit of membrane tension in the light-cone Hamiltonian of the
supermembrane compactified on $T^2$ \cite{RT}. The higher order parts
$H_1$ and $H_2$ represent the cubic and quartic complicated interactions.
In the large N limit the energy for the static background is given by
$H = P_{+} = 2RN(\pi R_1R_2)^2/{\a}^3$, where we use the light-cone
conventions. This value is also provided from (\ref{HR}) because $Tr$
can be identified with $N/(2\pi)^2\int_0^{2\pi}dq\int_0^{2\pi}dp$
in the large N limit, where the trace of quantum operators becomes the
phase-space integration of the classical functions. This identification
is confirmed by comparing $1/(2\pi)^2 \int dqdpe^{i\m\cdot\p}
e^{i\n\cdot\p} = \delta_{\m+\n,0}, \p = (p,q)$ with (\ref{J})
\cite{BFSS,KR}. Since $P_{+} = M^2/2P_{-}$ with the momentum $P_{-}$
quantized to be $N/R$, the membrane mass is given by
$M = 2\pi R_1R_2N/(\a)^{3/2}$. Due to the area of membrane $(2\pi)^2
R_1R_2N$ the membrane tension is evaluated as $T_2 = 1/\,2\pi(\a)^{3/2}$
\cite{BFSS,AB}.

Now let us return to the Lagrangian $L$ in (\ref{LT}). In the action
$S = \int dtL$ the factor $R$ can be absorbed by the definition of the
world volume time $t \rightarrow \sqrt{\a}t/\,2\pi R$ as well as the
rescaling $A_t \rightarrow 2\pi RA_t/\sqrt{\a}$. Simultaneously we arrive at
a Lorentz and gauge invariant form of the action
\beq
S = \int dtTr \left( \frac{(2\pi)^3(\a)^{3/2}}{4}F_{\mu\nu}^2 - \frac
{\pi(R_1R_2)^2}{(\a)^{5/2}} + \frac{\pi}{\sqrt{\a}}(D_{\mu}Z_I)^2 +
\frac{[Z_I,Z_J]^2}{8\pi(\a)^{5/2}} \right).
\label{ST}\eeq
Here the large N limit is taken so that the matrices described by the
quantum phase-space variables approach the ordinary functions that depend
on the classical phase-space ones. Then the trace identified with
$N/(2\pi)^2\,\int dqdp$ is further rewritten in terms of dual valuables
$(x_1,x_2)$ as
\beq
 \frac{1}{(2\pi)^2 \R \S} \int_0^{\sqrt{N} \hat{L_1}}dx_1
\int_0^{\sqrt{N} \hat{L_2}}dx_2,
\label{x}\eeq
where $\hat{R_r} = \a/R_r$ are radii of circles of the dual torus and
$\hat{L_r} = 2\pi\hat{R_r}$. In the large N limit the effective theory
becomes a U(1) gauge theory in the three-dimensional dual spacetime with
a field strength $F_{\mu\nu} = \partial_{\mu}A_{\nu} - \partial_{\nu}
A_{\mu} + 2\pi\R\S\{A_{\mu},A_{\nu}\}$ with $\{A_{\mu},A_{\nu}\} =
\epsilon^{rs}\;(\partial A_{\mu}/\partial x_{r})(\partial A_{\nu}/\partial
x_{s})$, which seems to include a finite correction term. Similarly
$D_{\mu}Z_I,\; [Z_I,Z_J]$ are expressed as ${\partial}_{\mu}Z_I + 2\pi \R \S
\{ A_{\mu},Z_I \}, 2\pi i\R \S \{ Z_I,Z_J \}$. The U(1) gauge coupling
constant can be read off from (\ref{ST}) and (\ref{x}) as
\beq
g^2 = \frac{\sqrt{\a}}{2\pi R_1R_2}
\label{gm}\eeq
and then has mass dimension $1/2$. Moreover the rescaling of integration
variables as $t \rightarrow \sqrt{N}t, x_r \rightarrow \sqrt{N}x_r$
together with $Z_I = (2\pi \R\S\sqrt{\a})^{1/2}\Phi_I$ makes the action
 (\ref{ST}) except for the second term into the canonical form
\beq
S = \sqrt{N}\int dt \int_0^{\hat{L_1}}dx_1 \int_0^{\hat{L_2}}dx_2
\left( \frac{1}{4g^2}F_{\mu\nu}^2 + \frac{1}{2}(D_{\mu}\Phi_I)^2 +
\frac{g^2}{4}[\Phi_I,\Phi_J]^2 \right),
\label{Sx}\eeq
where $F_{\mu\nu} = \partial_{\mu}A_{\nu} - \partial_{\nu}A_{\mu} +
2\pi\R\S/\sqrt{N}\,\{A_{\mu},A_{\nu}\}, D_{\mu}\Phi_I = \partial_{\mu}
\Phi_I + 2\pi\R\S/\sqrt{N}\,\{A_{\mu},\Phi_I\}$ and $[\Phi_I,\Phi_J] =
2\pi i\R\S/\sqrt{N} \,\{\Phi_I,\Phi_J\}$. It is seen that the correction
term in the U(1) field strength is suppressed in the large N limit.

If we take the large N limit for the Hamiltonian (\ref{HR}) before taking
the quantum trace and perform the integration over the classical phase
space $(p,q)$, that is, over the two coordinates parametrizing a
membrane, then we again have the Hamiltonian identical to (\ref{HRH})
with sin$\frac{\pi}{N}(\m \times \n)$ replaced by $\frac{\pi}{N}
(\m \times \n)$. Therefore in comparing two
expressions (\ref{HRH}) and (\ref{Sx}) we interpret (\ref{HRH}) as
the Hamiltonian describing the membrane wrapped around the
rectangular target-space torus, which means that in the separation (\ref{Y})
$U_1, U_2$ are regarded as winding modes and $A_1, A_2$ as oscillating
modes of the membrane. In this sense in the $(p,q)$ space we recognize
that it is better to represent $2\pi\a(A_1,A_2)$ by $(Z_1,Z_2)$.
Recently in Ref.\cite{WPP} the supermembrane with winding has been studied
in the light-cone gauge. The mode decomposition of the Hamiltonian
expressed in terms of the structure constants of the group of
area-preserving diffeomorphism is in a similar form to our
decomposition (\ref{HRH}). On the other hand the gauge invariant action
 (\ref{Sx}) shows a system in the dual $(x_1,x_2)$ space and $A_r$ can
be regarded as gauge potentials literally.

The background solution (\ref{U}) represents a single infinite membrane
and yields the U(1) gauge theory. Now we will consider a background
configuration of $N_2$ parallel infinite membranes which is described by
block-diagonal $N \times N$ matrices with $N =N_1N_2$
\begin{eqnarray}
U_1 =(-\sqrt{N_1}R_1p) \otimes 1,& & U_2 = \sqrt{N_1}R_2q \otimes 1,
\end{eqnarray}
and all others $U_I = 0, I = 3,\cdots,9$,
where $(p,q)$ are $N_1 \times N_1$ matrices with $[q,p] = 2\pi i/N_1$ and
$1$ is the unit $N_2 \times N_2$ matrix. Non-block-diagonal fluctuations are
expressed as
\begin{eqnarray}
Y_{ab}^1 &=& - \sqrt{N_1}R_1 p \delta_{ab} + 2\pi{\a}A_1(p,q)_{ab},
\nonumber\\ Y_{ab}^2 &=& \sqrt{N_1}R_2 q\delta_{ab} + 2\pi{\a}A_2(p,q)_{ab}
\end{eqnarray}
with $a,b = 1,\cdots,N_2$. In the same way as (\ref{AZ}) they are expanded
as $A_r(p,q)_{ab} = 1/2\pi\sqrt{\a}\,\sum_{\m}A_{ab}^r(\m)J_{\m}$.
The commutator between $A_1$ and $A_2$ is calculated in the large $N_1$
expansion \cite{ML} as
\begin{eqnarray}
(A_1*A_2 - A_2*A_1)_{ab} &=& [A_1,A_2]_{ab} + \sum_{k=1}^{\infty}
(\frac{\pi i}{N_1})^k\frac{1}{k!}\sum_{\m,\n}(\m \times \n)^k \nonumber\\
 & &  \times(A^1(\m)A^2(\n) - (-1)^kA^2(\n)A^1(\m))_{ab}J_{\m + \n}.
\end{eqnarray}
Using the dual coordinates in the leading order of the large $N_1$
expansion we have $[Y_1,Y_2]_{ab} = 2\pi iR_1R_2 \delta_{ab} +
(2\pi \a)^2i(F_{12})_{ab}$ where $F_{12}$ is the usual U($N_2$) gauge field
strength, and also $D_tY^r = 2\pi{\a}F_{tr},\, [Y^r,Y^I] = 2\pi{\a}iD_rZ_I$
with the standard U($N_2$) covariant derivative. In the large $N_1$ limit
the $Tr$ turns out to be $N_1/(2\pi)^2 \,\int dpdq tr$ where $tr$ is the
trace for $N_2 \times N_2$ matrices, so that we get a
three-dimensional U($N_2$) gauge theory. The suitable
rescalings discussed above yield the U($N_2$) gauge action
\beq
S = \sqrt{N_1} \int dt \int_0^{\hat{L_1}}dx_1 \int_0^{\hat{L_2}}dx_2
tr \left( \frac{1}{4g^2}F_{\mu\nu}^2 + \frac{1}{2}(D_{\mu}\Phi_I)^2 +
\frac{g^2}{4}[\Phi_I,\Phi_J]^2 \right),
\eeq
where $[\Phi_I,\Phi_J]$ is the usual U($N_2$) commutator and
the U($N_2$) gauge coupling constant  also scales as (\ref{gm}).
The last term has a favorite coefficient, since the further rescaling
$\Phi_I \rightarrow \Phi_I/g$ leads to $tr(F_{\mu\nu}^2/2 +
(D_{\mu}\Phi_I)^2 + [\Phi_I,\Phi_J]^2/2)\,/\,2g^2$ which is compared
with the starting Lagrangian (\ref{Lg}).

Let us next consider a four-brane solution expressed in a direct
tensor-product form
\begin{eqnarray}
U_1 = -\sqrt{n_1}R_1p_1\otimes1,& U_2 = \sqrt{n_1}R_2q_1\otimes1 ,
\nonumber\\ U_3 = -1\otimes\sqrt{n_2}R_3p_2,&
U_4 = 1\otimes\sqrt{n_2}R_4q_2 ,
\end{eqnarray}
and all others $U_I = 0, I = 5,\cdots,9$,
where $N = n_1n_2$ and $p_i, q_i$ are $n_i \times n_i$ matrices obeying
$[q_i,p_i] = 2\pi i/n_i, i = 1,2$. Defining the dual variables as
\beq
x_1 = \frac{\a\sqrt{n_1}}{R_1}q_1,\; x_2 = \frac{\a\sqrt{n_1}}{R_2}p_1,\;
x_3 = \frac{\a\sqrt{n_2}}{R_3}q_2,\; x_4 = \frac{\a\sqrt{n_2}}{R_4}p_2
\eeq
and substituting the fluctuations around the four-brane background
$Y^r = U_r + 2\pi{\a}A_r, r = 1,\cdots,4$ into the starting Lagrangian
(\ref{LR}) we obtain also the expressions (\ref{LT}) and (\ref{HR})
with $r = 1,\cdots,4$ and $I = 5,\cdots,9$ where $F_{12}^2$ and
$(R_1R_2)^2$ are replaced by $\sum_{r<s}F_{rs}^2$ and $(R_1R_2)^2 +
(R_3R_4)^2$ respectively. The fluctuations $A_r$ are also expanded as
\beq
A_r = \frac{1}{2\pi\sqrt{\a}}\sum_{\m,\n}A_{\m,\n}^rJ_{\m}(p_1,q_1)
\otimes J_{\n}(p_2,q_2).
\eeq
The comparison of the trace of quantum operators
\beq
Tr(J_{\m}(p_1,q_1)\otimes J_{\n}(p_2,q_2)\, J_{{\m}'}(p_1,q_1)\otimes
J_{\n'}(p_2,q_2)) = N\delta_{\m+\m',0}\delta_{\n+\n',0}
\eeq
with the four-dimensional phase-space integration
\beq
\frac{1}{(2\pi)^4} \int \prod_{i=1}^{2}dq_idp_i e^{i(\m\cdot\p_1 +
\n\cdot\p_2)}e^{i(\m'\cdot\p_1 + \n'\cdot\p_2)} = \delta_{\m+\m',0}
\delta_{\n+\n',0}, \;\p_i = (p_i,q_i)
\eeq
makes $Tr$ identified with $N/(2\pi)^4\,\int \prod_{i}dq_idp_i$
in the large N limit. The mass square of the static four-brane configuration
is obtained by $M^2 = 4{\pi}^2N^2((R_1R_2)^2 + (R_3R_4)^2)/\a^3$. Since the
area of the infinite four-brane is given by $n_1n_2\prod_{r}2\pi R_r$ the
four-brane tension is estimated as $T_4 = ((R_1R_2)^2 + (R_3R_4)^2)^{1/2}
/\,(2\pi)^3(\a)^{3/2}R_1R_2R_3R_4$. If we choose $R_1R_2 = R_3R_4$ to be
$\sqrt{2}\a$, $T_4$ becomes $1/\,(2\pi)^3(\a)^{5/2}$ which combines with
$T_2 = 1/\,2\pi(\a)^{3/2}$ to satisfy a relation $T_p = (4\pi^2\a)^{3-p}
T_{6-p}$ \cite{A,P,GHT} for the $p$-brane tensions $T_p$. Through the
rescaling $t \rightarrow \sqrt{\a}t/\,2\pi R$ a Lorentz invariant action is
again obtained to be in the same form as (\ref{ST}) up to the second term.
Then in the large N limit the $Tr$ becomes
\beq
N\int\prod_{i=1}^2 \frac{dq_i}{2\pi}\frac{dp_i}{2\pi} = \frac{1}
{(2\pi)^4\R\S\T\U} \int_{0}^{\sqrt{n_1}\hat{L_1}}dx_1 \int_{0}
^{\sqrt{n_1}\hat{L_2}}dx_2 \int_{0}^{\sqrt{n_2}\hat{L_3}}dx_3
\int_{0}^{\sqrt{n_2}\hat{L_4}}dx_4,
\eeq
where the size of the four cycles of the four-torus are $L_r = 2\pi\hat
{R}_r$ with $\hat{R}_r = \a/R_r$. In this way the
four-brane solution contained in the starting zero-brane theory leads to
the five-dimensional U(1) gauge theory whose coupling constant scales as
\beq
g^2 = \frac{2\pi(\a)^{5/2}}{R_1R_2R_3R_4}
\label{gf}\eeq
with mass dimension $-1/2$. The choice $n_1 = n_2 = \sqrt{N}$ and the
rescalings of $Z_I = ((2\pi)^3\R\S\T\U\sqrt{\a})^{1/2}\Phi_I$ as well as
the integration variables also provide a compact expression of the canonical
gauge action similar to (\ref{Sx}) with the five-dimensional  integrations
multiplied by $N^{3/4}$. It is interesting that the behaviors of $g^2$
inversely proportional to the volume of the torus in (\ref{gf})
and (\ref{gm}) are similar to the results of Susskind et al.
\cite{S,SS,FHRS} which are obtained by equating the energies
stored in the U(1) electric field or magnetic flux in the
SYM gauge theory in the dual torus with the corresponding ones
in the toroidally compactified M(atrix) theory.
It is, however, unclear in our framework how a new coordinate emerges for
the M(atrix) theory compactified on a two-torus when the torus shrinks
to zero size, which particular limit is considered to yield the
ten-dimensional type IIB string theory \cite{SS,FHRS}.

In conclusion from the D0-brane theory we have derived a sytem that is
interpreted in terms of the canonical phase-space variables as the
M(atrix) theory compactified on a two- or four-torus and alternatively
represented in terms of the dual phase-space variables as the three- or
five-dimensional gauge theory. In the former point of view the four-brane
tension with suitable dimension has been presented.
We compare our prescription with that of
Ganor et al. \cite{GRT,AT}, who replace by hand the matrices $X^{\mu}$ on
the $p$-dimensional torus with the covariant derivatives
$\nabla^{\mu} = i\frac{\partial}{\partial\hat{x}_{\mu}} + A^{\mu}(\hat{x})$
where $\hat{x}_{\mu}$ are added as the dual extra coordinates on the dual
torus. This replacement changes the Lagrangian of the U(N) $0+1$ dimensional
gauge theory into that of the U(N) $p+1$ dimensional gauge theory where the
dual space integration is added by hand. In their transformed Lagrangian
there is no term corresponding to the second term in (\ref{ST}) of our
Lorentz invariant Lagrangian.

Recently in Ref.\cite{VK} starting from the M(atrix) theory the effective
gauge theories on the $p$-brane world volumes have been derived without
using T-duality by replacement of covariant derivatives which are interpreted
to be expressed in terms of the coordinates parametrizing the $p$-brane
rather than the dual coordinates. In their shuffles of matrix fields the
linear terms in the gauge potential remain in the effective Lagrangians.

In our approach we have demonstrated that the suitable extractions of the
dual coordinates from the basic canonical phase-space variables
parametrizing the membrane and four-brane solutions lead to appearances
of the dual space intergrations. We have seen that the dual coordinates
together with the D2- and D4-brane solutions emerge from only the degrees
of freedom of D0-brane, where the trace in the starting Lagrangian
contains the dual space integration in a natural way. In the dual spacetime
the appropriate rescalings yield the Lorentz and gauge invariant action in a
favorite form with the gauge coupling constant characterized by the volume
of the dual torus. We expect that our demonstrations will be useful to get
a better understanding of the toroidal compactification
in the M(atrix) theory.

\end{document}